\documentclass[twocolumn,letter]{jpsj3}
\usepackage{amssymb, amsmath}
\usepackage{txfonts}
\usepackage{graphics}
\usepackage{epsfig}
\usepackage{times}
\usepackage{dcolumn}
\usepackage{bm}
\usepackage{cite}
\usepackage{ulem}
\usepackage{braket}
\usepackage{color}
\usepackage{mathrsfs}

\title{Finite-temperature properties of excitonic condensation in the extended Falicov-Kimball model:
Cluster mean-field-theory approach}

\author{
Masahiro Kadosawa$^1$\thanks{afna1728@chiba-u.jp}, 
Satoshi Nishimoto$^{2,3}$, 
Koudai Sugimoto$^4$, 
Yukinori Ohta$^1$}
\inst{$^1$Department of Physics, Chiba University, Chiba 263-8522, Japan\\
$^2$Department of Physics, Technical University Dresden, 01069 Dresden, Germany\\
$^3$Institute for Theoretical Solid State Physics, IFW Dresden, 01069 Dresden, Germany\\
$^4$Department of Physics, Keio University, Yokohama 223-8522, Japan}

\date{\today}

\abst{
We study the electron-hole pair (or excitonic) condensation in the extended Falicov-Kimball 
model at finite temperatures based on the cluster mean-field-theory approach, where we 
make the grand canonical exact-diagonalization analysis of small clusters using the sine-square 
deformation function.  We thus calculate the ground-state and finite-temperature phase 
diagrams of the model, as well as its optical conductivity and single-particle spectra, thereby 
clarifying how the preformed pair states appear in the strong-coupling regime of excitonic 
insulators. We compare our results with experiments on Ta$_2$NiSe$_5$.  
}

\begin{document}

\maketitle


The electron-hole pair (or excitonic) condensation \cite{mott,jerome,halperin} in transition-metal 
 chalcogenides and oxides has attracted much attention in recent years \cite{kunes,ssp}.  One 
of the representative materials is Ta$_2$NiSe$_5$ \cite{wakisaka,wakisaka2,kaneko,seki1}, where 
it was pointed out that the system is a spin-singlet excitonic insulator (EI) in the strong-coupling 
regime, so that the conventional phase diagram \cite{kozlov} breaks down \cite{sugimoto}; 
i.e., even though the noninteracting band structure is semimetallic, the system above the 
transition temperature ($T_c$) is not a semimetal, but rather a state of strongly coupled 
preformed pairs with a finite band gap.  
A novel insulator state exhibiting a variety of intriguing physical properties is thus 
expected to occur.  However, not much is known about the preformed pair states in 
EI models, among which the simplest spinless fermion model for spin-singlet excitonic 
condensation is the extended Falicov-Kimball model (EFKM) \cite{zenker,seki2}.  

In this paper, we study finite-temperature properties of the EFKM at half filling using 
the cluster mean-field-theory (CMFT) approach 
\cite{shibata,albuquerque,brzezicki,suzuki,gotfryd,singhania}, 
whereby we can take into account the quantum (as well as thermal) fluctuations of the 
system, allowing for any finite-temperature phase transition. 
We make a grand canonical exact-diagonalization analysis of small clusters, employing 
the so-called sine-square deformation (SSD) function \cite{hotta1,hotta2}. Thus, we 
calculate a number of physical quantities in the semi-thermodynamic limit, which can be 
approached by a small-cluster analysis.  

In what follows, we will first present the model and method of our calculations; in particular, 
we discuss the CMFT approach using the grand canonical exact-diagonalization analysis 
of small clusters in some detail, where we use the SSD function.  We will then show our 
results for the ground-state and finite-temperature phase diagrams of the model.  
We will also show temperature dependence of the optical conductivity and single-particle 
spectra, thereby clarifying how the preformed pair states appear in the strong-coupling 
regime of the EIs. Finally, we will compare our results with experiments on Ta$_2$NiSe$_5$ 
and discuss implications of our results to the strong-coupling nature of the EI states. 

We thus shed some light on the finite-temperature properties, or the preformed pair states 
above $T_c$, of the strong-coupling regime of the EIs for the first time, the clarification 
of which has long been sought for \cite{sugimoto} but, to the best of our knowledge, 
has not been fully discussed so far. 

\begin{figure}[thb]
\centering
\includegraphics[width=0.65\linewidth]{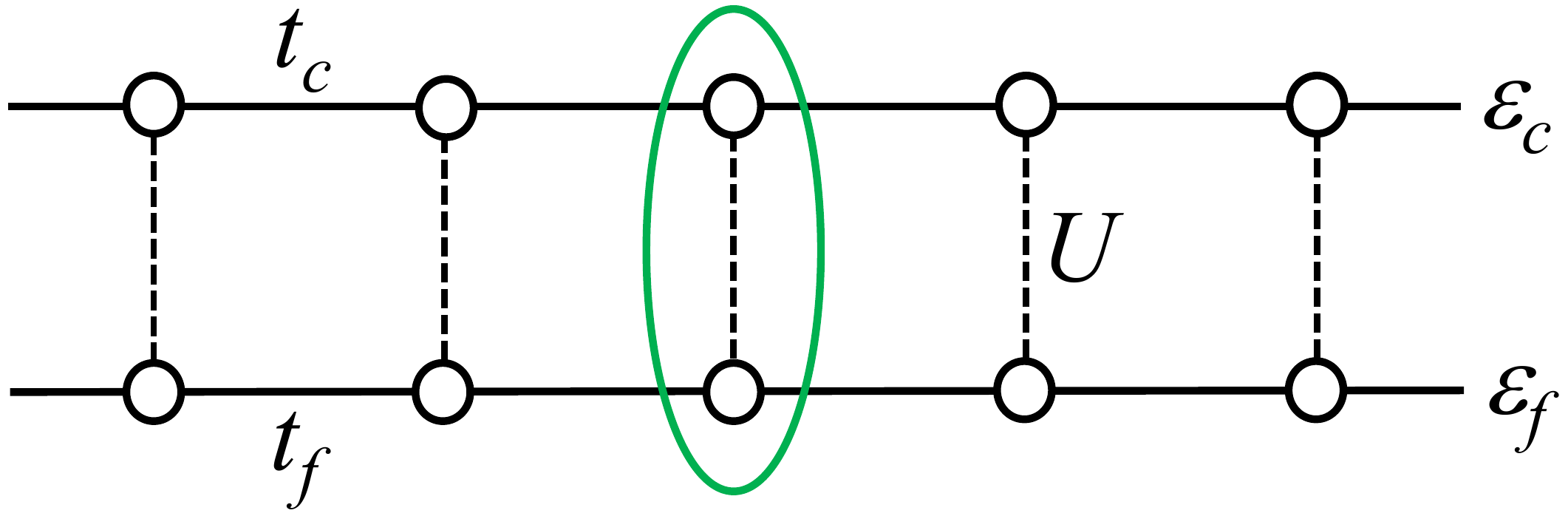}
\caption{(Color online) 
Schematic representation of the 1D EFKM.  The mean-field bond is indicated by an ellipse.  
}\label{fig1}
\end{figure}



A minimal theoretical model for the spin-singlet EI states, such as in Ta$_2$NiSe$_5$, 
is the Falicov-Kimball model \cite{FKM} extended by including a finite width of the valence 
band, which is referred to as the EFKM \cite{zenker,seki2}.  The Hamiltonian reads 
\begin{align}
\nonumber
	\cal{H} &= -t_c\sum_{\langle i,j \rangle}c_i^\dagger c_j-t_f\sum_{\langle i,j \rangle}f_i^\dagger f_j
+U\sum_i c_i^\dagger c_i f_i^\dagger f_i\\
	&+(\varepsilon_c-\mu)\sum_i c_i^\dagger c_i+(\varepsilon_f-\mu)\sum_i f_i^\dagger f_i,
	\label{ham1}
\end{align}
where $c_i$ and $f_i$ are annihilation operators of spinless fermions (which is referred 
to as an electron hereafter) in the conduction-band ($c$) and valence-band ($f$) orbitals 
at site $i$, respectively.  
We define the energy-level splitting between the $c$ and $f$ orbitals as 
$D=\varepsilon_c-\varepsilon_f$ $(>0)$, and the hopping parameters in the respective 
orbitals as $t_c$ and $t_f$.  $\mu$ is the chemical potential.  
Regarding the modeling of Ta$_2$NiSe$_5$, we consider a one-dimensional (1D) lattice 
under the implicit assumption of the presence of weak three-dimensionality 
{(see below)}, representing a direct band-gap semiconductor (or a semimetal) 
with $t_c$ $(>0)$ and $t_f$ $(<0)$, where the direct hopping of electrons between the $c$ 
and $f$ orbitals is prohibited (see Fig.~\ref{fig1}).  
We assume the on-site Coulomb repulsion between electrons to be $U$, which acts 
as the on-site attractive interaction between an electron and a hole.  
We restrict ourselves to the case at half-filling, so that we have either a 
semiconductor at $D>2(t_c+|t_f|$) or a semimetal at $D<2(t_c+|t_f|$) when $U=0$.  
Hereafter, we assume $t_c=1$ (unit of energy) and $t_f=-0.3$, unless otherwise indicated.  

Here, we note that, since the direct hopping of electrons between the $c$ and $f$ orbitals 
is prohibited in this model, the operators of the total number of electrons in each orbital have 
a simultaneous eigenstate of the Hamiltonian, so that any physical quantity changes 
discontinuously (due to discontinuous change in the total number of electrons in each 
orbital) when calculated, e.g., as a function of the parameters involved in the Hamiltonian.  
In small-cluster calculations, this situation leads to an apparently unphysical parameter 
dependence of the calculated physical quantity.  However, we will show below that this 
difficulty may essentially be suppressed by introducing the SSD function to our 
exact-diagonalization calculations of small clusters.  


We employ the CMFT to study phase transitions emerging in the system.  
In the CMFT, since only a part of the cluster (a site and/or a bond) is replaced 
by a mean field, quantum (as well as thermal) fluctuations within the cluster size 
can be taken into account \cite{CMFT}.  
The phase transition is then detected directly by a nonzero value of the mean field, 
which is regarded as the order parameter of the phase transition, as in the 
conventional mean-field theory.  
A recent example is the application of the CMFT to the 1D Heisenberg model 
for discussing the finite-temperature phase transition \cite{singhania,mermin}, 
where a customary assumption is adopted that the weak three-dimensional interchain 
coupling is implicitly introduced via the mean-field in the CMFT calculation, 
just as in our present calculation.  

We note here that the finite-temperature phase transition can be obtained in the 
conventional mean-field theory, but the effects of the fluctuations above $T_c$ 
cannot be seen.  In the CMFT, however, we can obtain the finite-temperature 
phase transition and at the same time the effects of the fluctuations above $T_c$ 
can be observed, as we will see below.  We also note that the density-matrix 
renormalization group (DMRG) calculation of the 1D EFKM has provided a successful 
results at zero temperature \cite{ejima}, which is however not suited for 
finite-temperature calculations of the present model \cite{schollwock}.  

We use finite-size clusters with $L \times 2$ orbitals, where the length $L$ is 
fixed to be odd.  Then, the Coulomb term at the center of the system 
is replaced by a mean-field bond (see Fig.~\ref{fig1}), namely,
\begin{align}
\nonumber
c_i^\dagger c_i f_i^\dagger f_i \simeq \langle c_i^\dagger c_i \rangle f_i^\dagger f_i 
&+ \langle f_i^\dagger f_i \rangle c_i^\dagger c_i \\
&- \langle f_i^\dagger c_i \rangle c_i^\dagger f_i 
- \langle c_i^\dagger f_i \rangle f_i^\dagger c_i,
\label{MFA}
\end{align}
where $\langle \cdots \rangle$ denotes the expectation value with respect 
to the ground state of the system at temperature $T=0$, while at finite $T$, 
it denotes the canonical average of an operator $A$ defined as 
$\langle A \rangle = \mathrm{Tr} \,A \exp(-\beta\mathcal{H}) /Z$, 
$Z=\mathrm{Tr} \exp(-\beta\mathcal{H})$, and $\beta=1/T$.  
Here, the Hamiltonian including the mean-field bond is 
solved numerically by a full exact-diagonalization of the cluster.  
Thus, the mean-field values 
$\langle c_i^\dagger c_i \rangle$, 
$\langle f_i^\dagger f_i \rangle$, 
$\langle f_i^\dagger c_i \rangle$, and 
$\langle c_i^\dagger f_i \rangle$ 
are calculated self-consistently. 
The chemical potential $\mu$ is also determined so as to fulfill the condition 
$\langle c_i^\dagger c_i \rangle+\langle f_i^\dagger f_i \rangle=1$.


Now, let us compute the mean-field values defined above.  
Since they are local quantities, we may approximately obtain them in the bulk 
limit even within the use of small clusters by applying the `grand canonical' 
analysis.  In this analysis, the original Hamiltonian consisting of local terms 
$\mathcal{H}_i$, defined as in Ref.~\cite{hotta1}, is deformed as 
\begin{equation}
{\mathcal H}_{\rm deform}= \sum_{i=1}^{L} {\mathcal H}_i f(\bm r_i),
\label{defham}
\end{equation}
where $f(\bm r)$ is an externally given function, which varies smoothly 
from the maximum at the center of the cluster [$i=(L+1)/2$] to zero at 
the edges of the cluster.  For such a function, we typically adopt the 
so-called sine-square deformation (SSD) function, which provides a 
smooth boundary condition.  For the 1D system, the SSD function is given as 
\begin{equation}
f_{\rm SSD}(i)=\sin^2\left(\frac{\pi}{L}\big(i-\frac{1}{2}\big)\right) 
\label{fssd1d}
\end{equation}
with either $i=1,\cdots,L$ for the on-site and $U$ terms or 
$i=3/2,\cdots,L-1/2$ for the hopping terms between sites $i$ and $i+1$.  
This deformation spatially scales down the energy from unity at the 
system center toward zero at the open edge sites, which introduces 
the renormalization of the energy levels in a way reminiscent of Wilson's 
numerical renormalization group \cite{hotta2,okunishi}.  
As a consequence, the local quantities 
around the system center are self-organized to tune the particle 
number of the bulk states to their thermodynamic limit by using the edges 
as a reservoir.  In this way, our grand canonical analysis using SSD optimally 
realizes the bulk eigenstate basis at the center of a small cluster.  
Smooth variations of the physical quantities calculated as a function of 
temperature, as well as of the internal parameters of the EFKM, are thereby 
demonstrated, as we will show below.  

\begin{figure}[thb]
\centering
\includegraphics[width=0.87\linewidth]{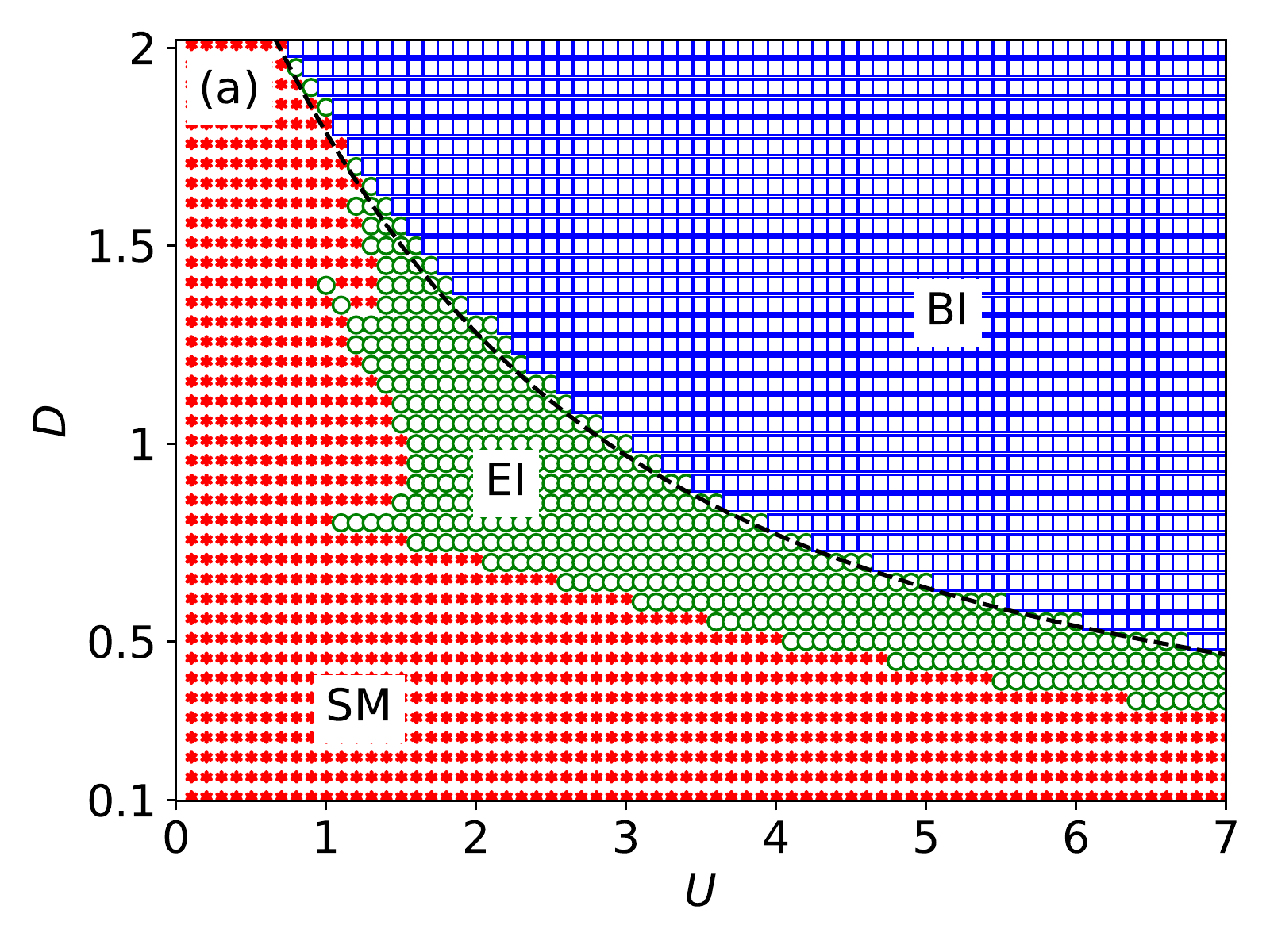}
\includegraphics[width=0.87\linewidth]{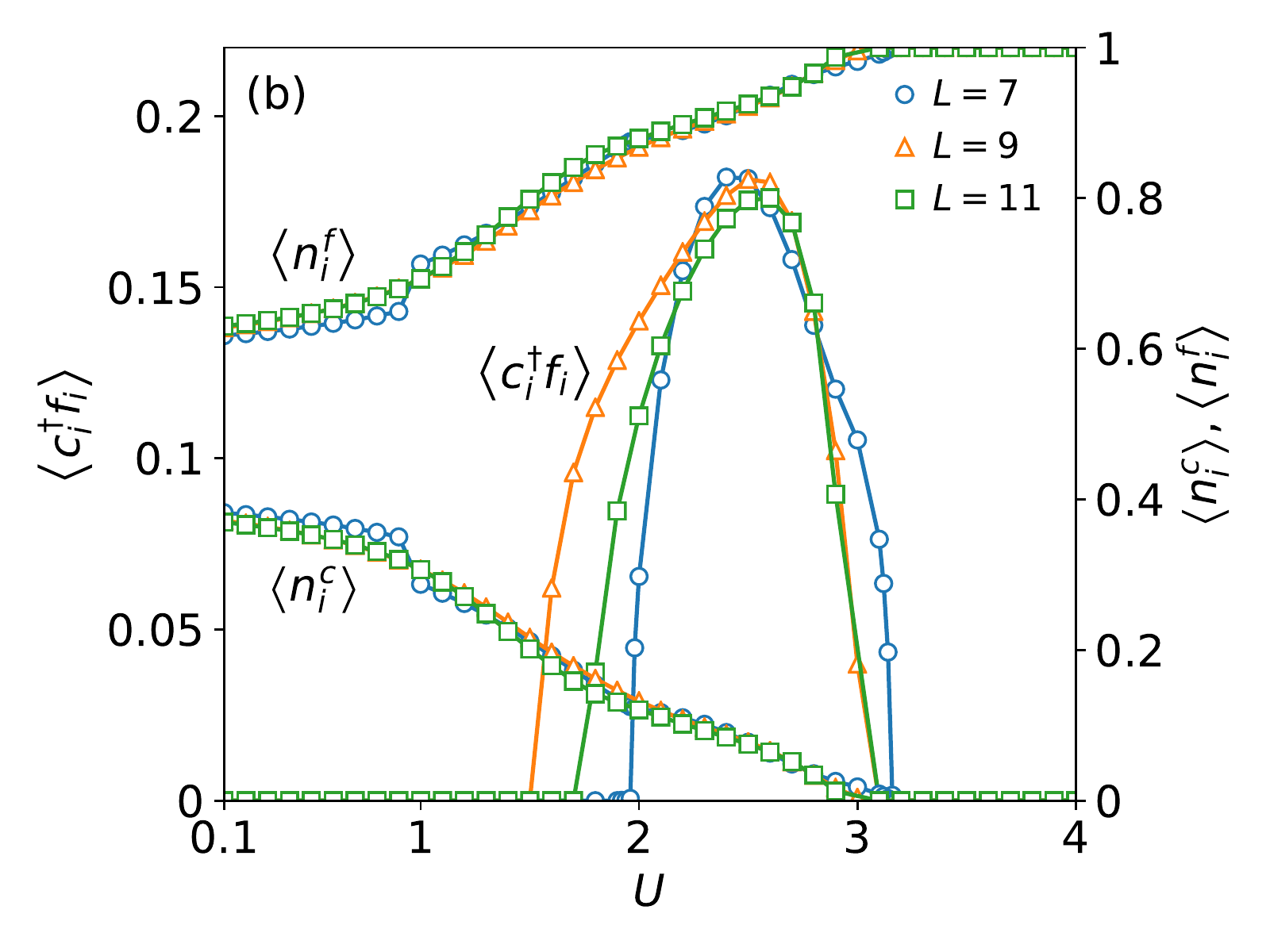}
\caption{(Color online) 
(a) Calculated ground-state phase diagram of the EFKM 
on the $(U,D)$ plane, which contains the BI (blue), EI (green), and normal 
SM (red) phases.  The dashed line indicates the exact analytical solution of 
the BI-EI phase boundary. The cluster of $L=9$ is used. 
(b) Calculated order parameter $\langle c_i^\dagger f_i \rangle$ 
and numbers of electrons in the $c$ and $f$ orbitals 
$\langle n_i^c\rangle$ and $\langle n_i^f\rangle$ as a function of $U$. 
$D=1$ is assumed. We compare the results obtained for the clusters 
of $L=7$ (blue), $L=9$ (orange), and $L=11$ (green).  
}\label{fig2}
\end{figure}



First, let us discuss the ground state of the EFKM at $T=0$ K.  
The calculated ground-state phase diagram of the model is shown 
in Fig.~\ref{fig2}(a) on the $(U,D)$ plane, where we find that the 
excitonic insulator (EI) phase actually occurs between the band 
insulator (BI) and normal semimetallic (SM) phases.  
The calculated phase boundary between the BI and EI phases agrees 
well with the exact BI-EI phase boundary 
$D=\sqrt{4(t_c+|t_f|)^2+U^2}-U$ obtained analytically 
\cite{kocharian} and indicated by the dashed line.  Here, we do not 
take into account the staggered orbital order phase, which should 
appear around $D=0$ \cite{ejima}.  

We note that, unlike in the numerically exact DMRG solution given in 
Ref.~\cite{ejima}, the EI phase appears only near the BI-EI phase boundary. 
This is because the EI phase cannot acquire sufficient energy-gain 
in the small $U$ region in the present CMFT calculations.  In this region, 
the energy-gain in the EI-state formation is exponentially small.  
We then suggest that the absence of the EI phase at small $U$ is an 
artifact of the SSD because some uncertainty in the electron number 
is always involved when applying the SSD \cite{hotta2}, just as the 
Mott insulating state immediately destabilized away from half-filling 
at small $U$ even within the mean-field level. 

We also note that the phase boundary between the EI and SM phases 
exhibits a nonmonotonous curve, which may be due to the 
above-discussed discontinuous behavior of small clusters of the EFKM.  
However, we should emphasize that the CMFT calculation indeed 
successfully provides the continuous EI phase between the BI and 
SM phases with the help of the SSD function.  


The calculated order parameter $\langle c_i^\dagger f_i \rangle$ 
and numbers of electrons in the $c$ and $f$ orbitals 
$\langle n_i^c\rangle=\langle c_i^\dagger c_i \rangle$ and 
$\langle n_i^f\rangle=\langle f_i^\dagger f_i \rangle$ 
are shown in Fig.~\ref{fig2}(b) as a function of $U$ at $D=1$ for 
the clusters of $L=7$, $9$, and $11$.  We find that the cluster-size 
dependence of the results, even for the order parameter 
$\langle c_i^\dagger f_i \rangle$, is not very strong.  
We also find that the numbers of electrons $\langle n_i^c\rangle$ 
and $\langle n_i^f\rangle$ vary smoothly as a function of $U$ owing 
to the SSD function, although we still notice a small discontinuity 
at $U\simeq 1$ for the $L=7$ cluster.  
We find again that $\langle c_i^\dagger f_i \rangle$ vanishes rather 
rapidly when $U$ is small, as discussed above.  


\begin{figure}[thb]
\centering
\includegraphics[width=0.87\linewidth]{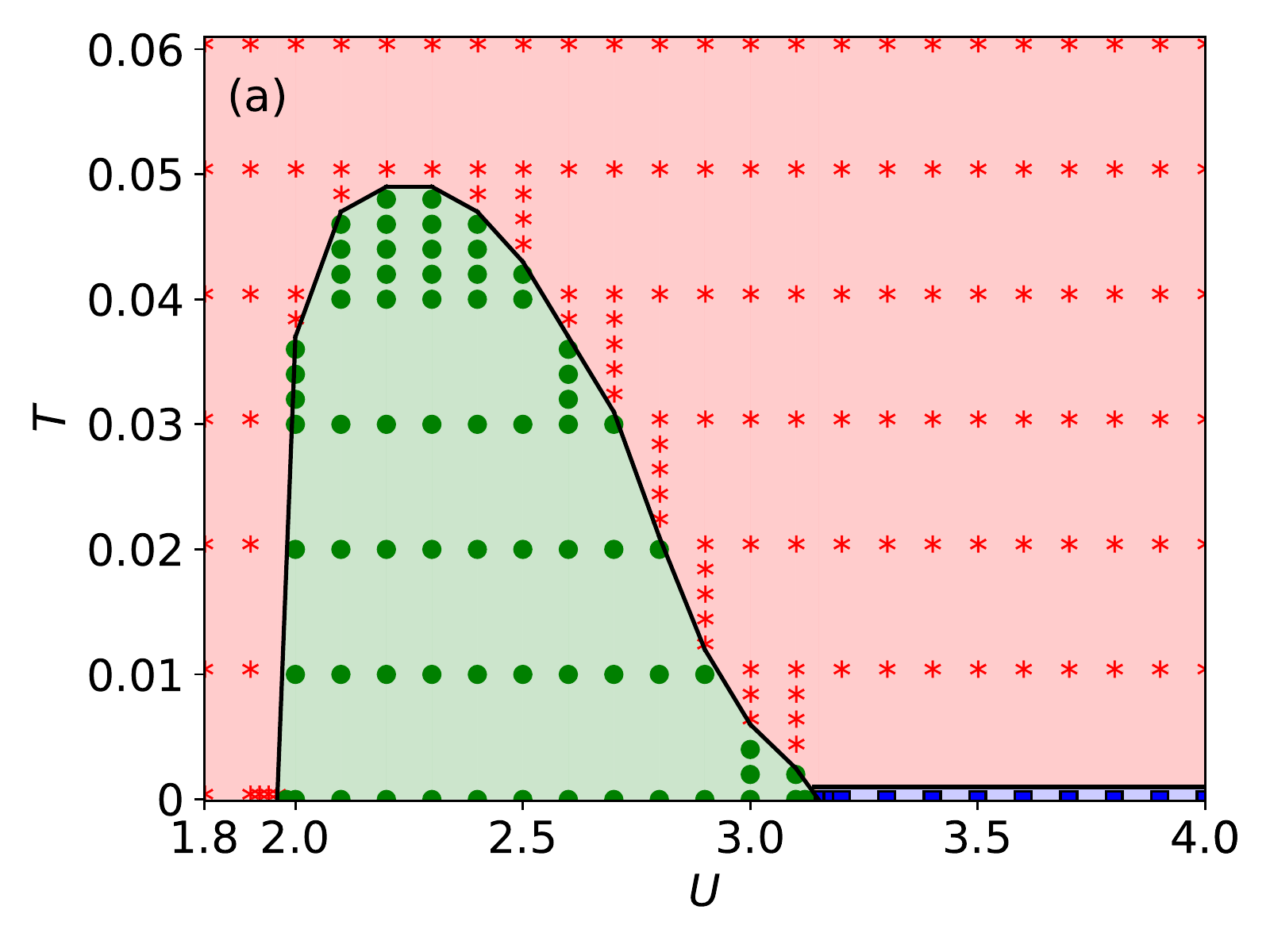}
\includegraphics[width=0.87\linewidth]{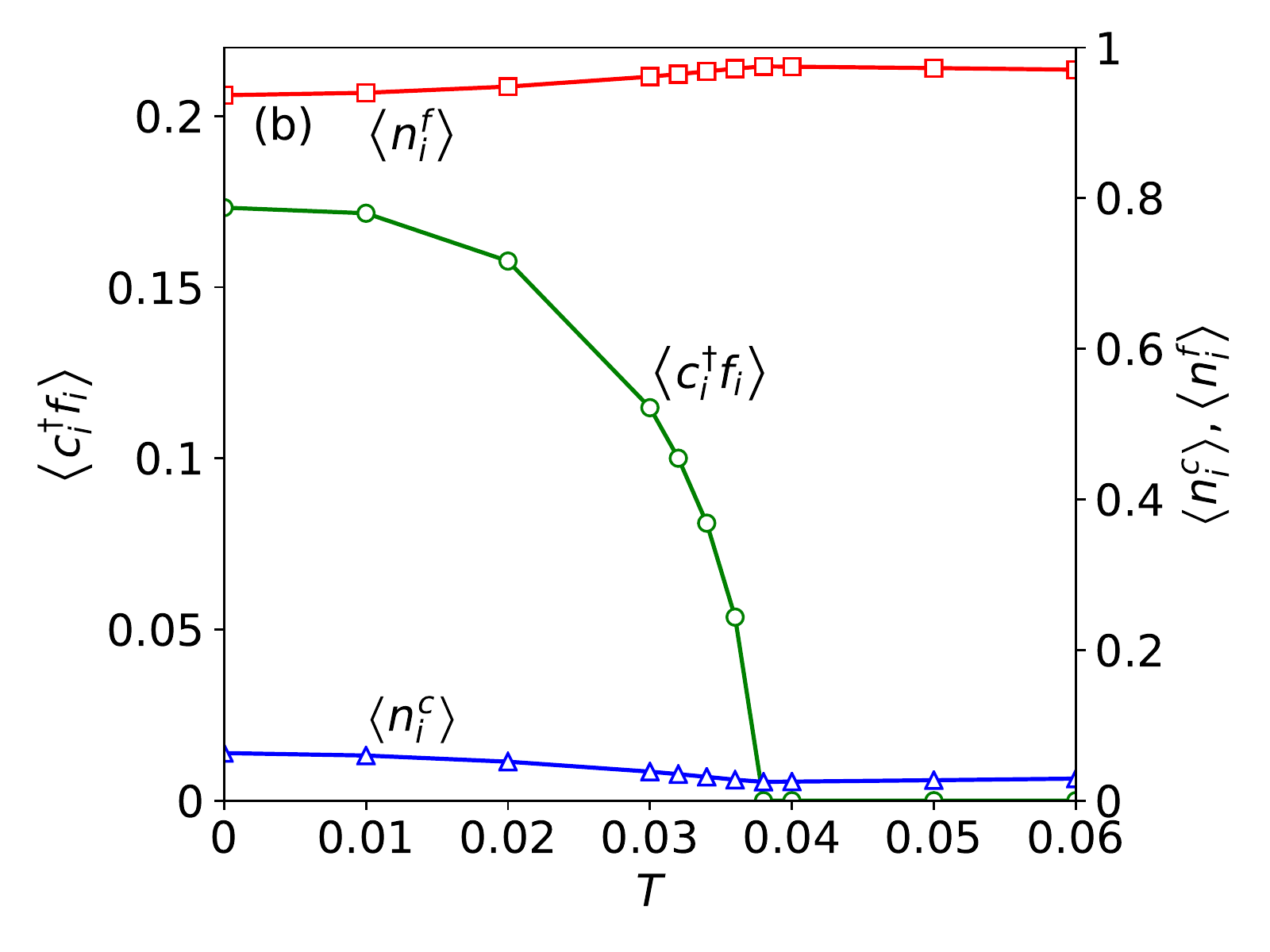}
\caption{(Color online) 
(a) Calculated finite-temperature phase diagram of the EFKM as a function 
of $U$, where the region $\langle c_i^\dagger f_i \rangle \ne 0$ is indicated by 
the green circles.  We use the cluster of $L=7$ and assume $D=1$.  
(b) Calculated temperature dependence of the order parameter 
$\langle c_i^\dagger f_i \rangle$ and numbers of electrons in the $c$ and $f$ 
orbitals $\langle n_i^c\rangle$ and $\langle n_i^f\rangle$.  
We use the cluster of $L=7$ and assume $U=2.6$ and $D=1$.  
}\label{fig3}
\end{figure}

Next, let us discuss the finite-temperature phase diagram of the 
EFKM.  The calculated result is shown in Fig.~\ref{fig3}(a) as a 
function of $U$ at $D=1$.  We find that a dome-like shape of the EI phase 
actually occurs as a function of $U$ at $2\lesssim U\lesssim 3.2$, which is 
between the BI (at $3.2\lesssim U$) and SM (at $U\lesssim 2$) phases.  Thus, 
the finite-temperature phase transition actually occurs at $T=T_c$, 
where the value of $T_c$ is found to be rather low.  
Although we may point out on the one hand that the finite-temperature 
phase transition does not occur in pure 1D systems due to thermal and 
quantum fluctuations, we may argue on the other hand that any 
mean-field-type treatment of quantum systems should provide a 
finite-temperature phase transition, the $T_c$ of which may well be low 
if partial inclusion of quantum fluctuations is made by, e.g., the CMFT 
as in the present case.  
The obtained value of $T_c$, therefore, does not have a quantitative 
significance unless the weak three-dimensionality in the real system is 
introduced explicitly via the realistic interchain coupling parameters, 
thereby making a quantitative calculation.  We should, however, note 
that the properties of the model above $T_c$ can be discussed in the 
framework of the present CMFT approach, as we will see below.  

The calculated temperature dependence of the order parameter 
and numbers of electrons in the $c$ and $f$ orbitals is shown 
in Fig.~\ref{fig3}(b).  We find that the system undergoes a continuous 
(or second-order) phase transition, as is evident in the behavior of 
$\langle c_i^\dagger f_i \rangle$.  We also find that the value of 
$\langle n_i^c\rangle$ ($\langle n_i^f\rangle$) increases (decreases) 
with decreasing temperature below $T_c$ due to the excitonic condensation 
(or spontaneous $c$-$f$ hybridization) although the change across 
the phase transition is rather small.  

\begin{figure}[thb]
\centering
\includegraphics[width=0.9\linewidth]{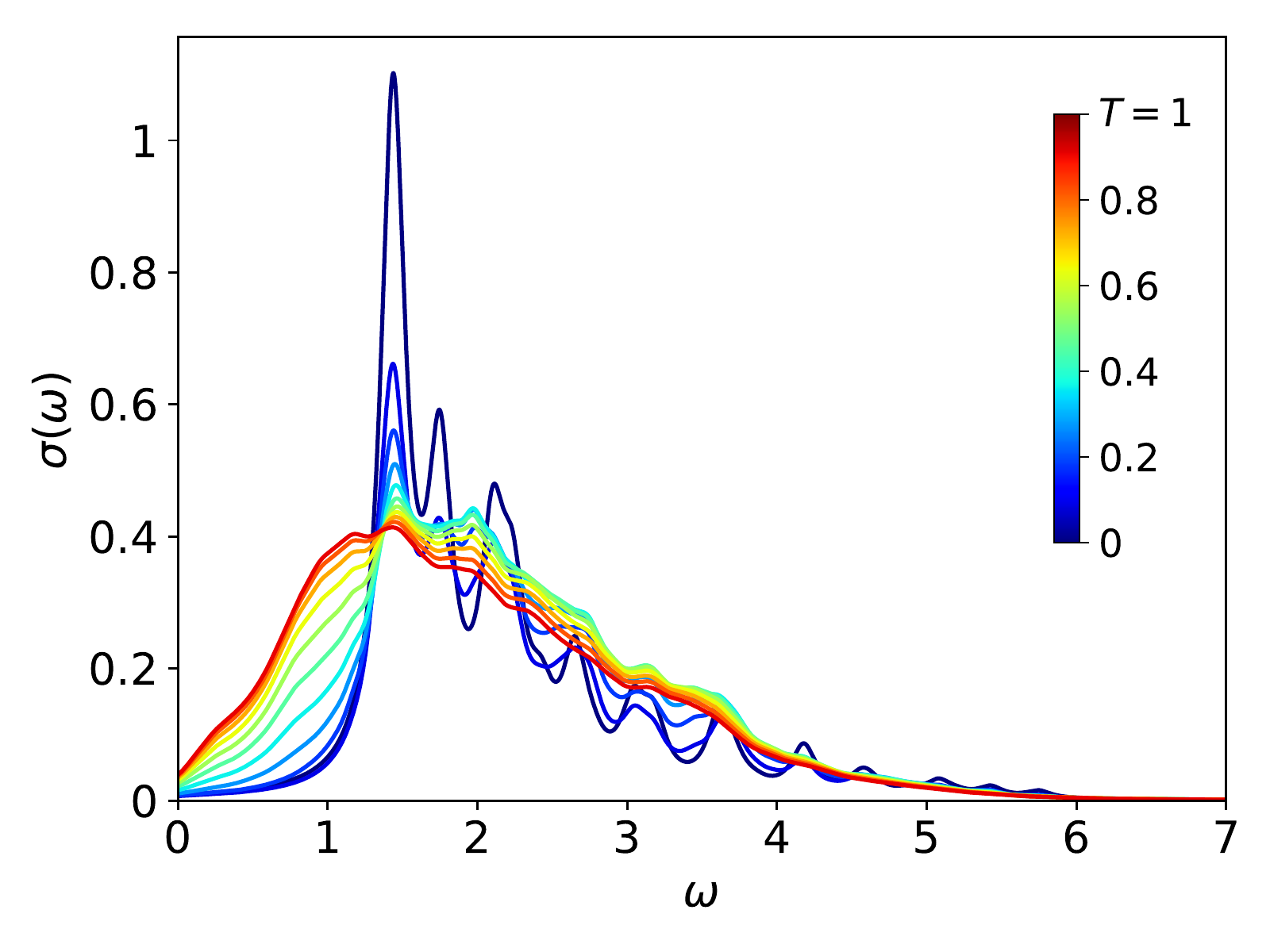}
\caption{(Color online) 
Calculated temperature dependence of the optical conductivity spectrum 
$\sigma(\omega)$. We use the cluster of $L=7$ and assume $U=2.5$ and 
$D=1$. The broadening parameter of the spectra is set to $\eta=0.1$.  
}\label{fig4}
\end{figure}


Finally, let us discuss the temperature dependence of the excitation 
spectra of the EFKM; in particular, we calculate the optical conductivity 
and single-particle spectra, which we will compare with experiments on 
Ta$_2$NiSe$_5$.  
The optical conductivity spectrum $\sigma(\omega)$ may be defined as 
\begin{align}
\nonumber
&\sigma(\omega) = \omega\, (1-e^{-\beta \omega}) \,I(\omega), \\
&I(\omega)=\frac{1}{\pi}\mathrm{Im}\,Z^{-1}\sum_{m,n} e^{-\beta E_m} \frac{\big| \langle \Psi_m \big| \,d\, \big| \Psi_n \rangle \big|^2}
{\omega-i\eta + E_m - E_n}
\label{opt}
\end{align}
with the dipole operator $d=-e\sum_{l=1}^L l\,\big(c_l^\dagger c_l + f_l^\dagger f_l -1 \big)$, 
where $E_m$ and $\Psi_m$ are the eigenvalue and eigenstate of the Hamiltonian, 
respectively \cite{gebhard}.  The single-particle spectrum may similarly be defined 
by replacing $d$ in Eq.~(\ref{opt}) with either $c_k$ ($f_k$) or $c_k^\dagger$ ($f_k^\dagger$) 
with momentum $k$, thereby simulating the angle-resolved photoemission or inverse 
photoemission spectrum $A(k,\omega)$. Below, we choose $k$ as a central site 
of the cluster in real space, to calculate the angle-integrated spectrum.  

The calculated results for the optical conductivity spectrum $\sigma(\omega)$ is 
shown in Fig.~\ref{fig4}, which may be compared with experiment for Ta$_2$NiSe$_5$; 
see Fig.~3 of Ref.~\cite{lu} and Fig.~2(c) of Ref.~\cite{larkin}.  
%
%
In our previous paper \cite{sugimoto}, we have confirmed that the main peak at 
$\omega\simeq 0.4$ eV in the observed optical conductivity spectrum originates 
from the repulsive interaction between electrons on the $c$ and $f$ orbitals at the 
same site, which is nothing but the attractive interaction between an electron on 
the $c$ orbital and a hole on the $f$ orbital at the same site.  In other words, the 
main peak in the optical conductivity spectrum is caused by the electron-hole pair 
formation. Here, we moreover confirm that, in both theory and experiments \cite{lu,larkin}, 
the main peak remains robust even above $T_c$, where the pairs are not condensed. 
Then, it seems quite natural to assume that the main peak of the optical conductivity 
reflects the preformed pair states of the system.  

We find that the temperature-induced spectral weight transfer, observed experimentally 
in Ta$_2$NiSe$_5$ \cite{lu,larkin}, is qualitatively well reproduced by our calculation; 
i.e., the spectral weight is transferred from high-frequency to low-frequency regions 
by increasing temperature.  
We should note that the change in the spectral features at $T_c$ is unnoticeably 
small, which is also consistent with experiments, where virtually no discontinuous 
changes occur at $T_c$ \cite{lu,larkin}.  The behavior of this peak thus illustrates 
the preformed electron-hole pair state in Ta$_2$NiSe$_5$, which appears even far 
above $T_c$.  

\begin{figure}[thb]
\centering
\includegraphics[width=0.8\linewidth]{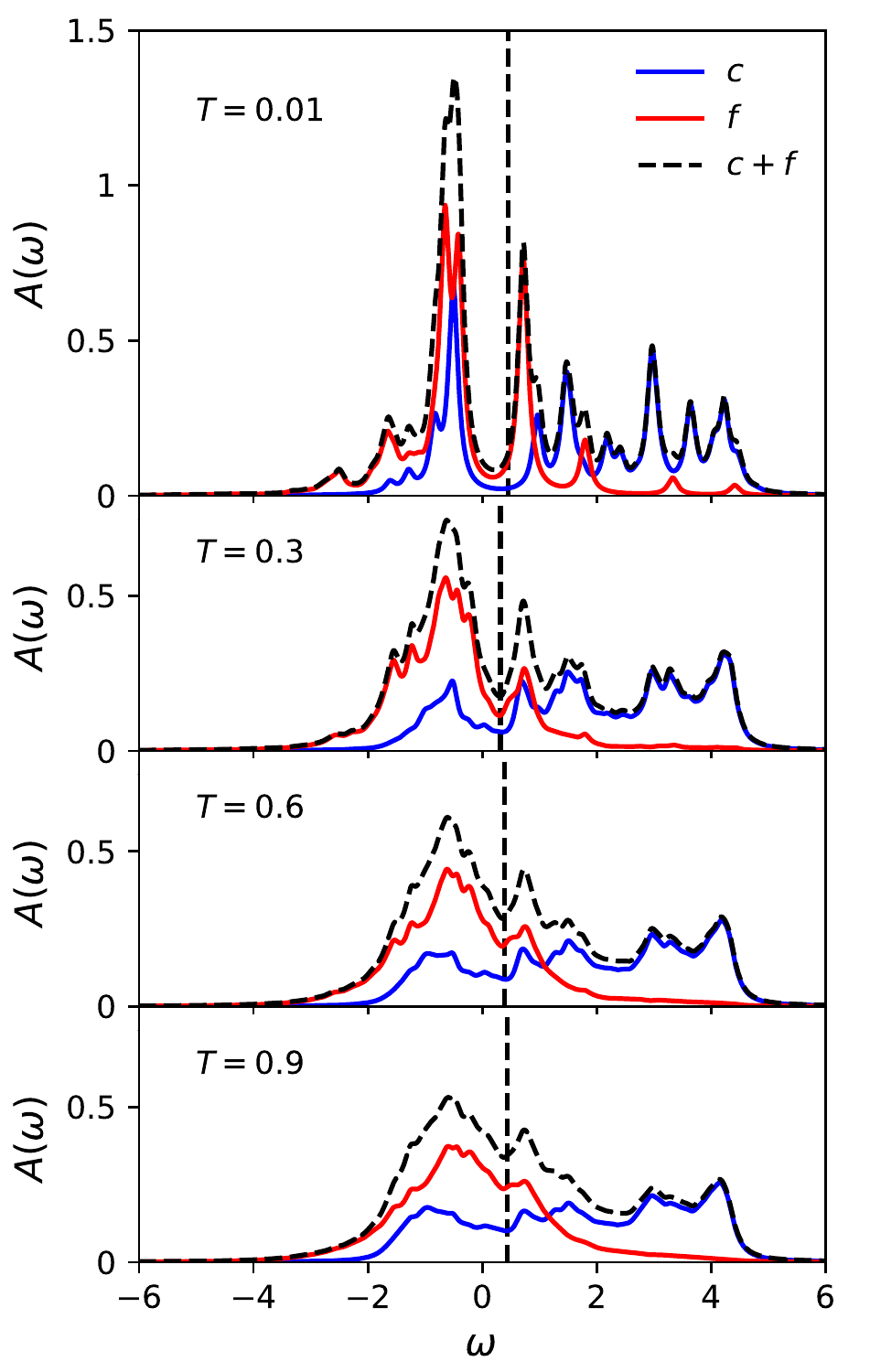}
\caption{(Color online) 
Calculated temperature ($T$) dependence of the $k$-integrated single-particle 
spectra $A(\omega)$, where $c$ and $f$ contributions are shown separately., 
together with the total spectral weight $c+f$.  The vertical line indicates the 
Fermi level.  We use the cluster of $L=7$ and assume $U=2.5$ and $D=1$. 
The broadening parameter of the spectra is set to $\eta=0.1$.  
}\label{fig5}
\end{figure}

The calculated results for the $k$-integrated single-particle spectrum $A(\omega)$ 
are shown in Fig.~\ref{fig5}, where we find that the band-gap feature observed at 
$T=0$ essentially remains even far above $T_c$ as a pseudogap-like structure, 
indicating that the electron-hole pairs survive robustly.  The temperature dependence 
of the angle-resolved photoemission spectra observed experimentally \cite{seki1} are 
consistent with our calculated results.  

Thus, the preformed pair state in the strong-coupling regime of excitonic insulators 
manifests itself in both the optical conductivity and single-particle spectra.  


In summary, we studied the excitonic condensation in the 1D EFKM 
at finite temperature based on the CMFT approach.  We obtained the 
ground-state and finite-temperature phase diagrams of the model using 
the grand canonical exact-diagonalization analysis of small clusters with 
the SSD function, whereby the unphysical temperature and parameter 
dependence of the results was suppressed.  We also presented the 
temperature dependence of the optical conductivity and single-particle 
spectra of the model and compared them with experiments on Ta$_2$NiSe$_5$.  
We thus discussed how the preformed pair state appears in the strong-coupling 
regime of the EI.  
We hope that more quantitative analyses of the experimental data will 
be made in future based on more realistic models \cite{sugimoto2,mazza} 
and more powerful computational techniques,\cite{seki3,nishida} 
to reveal the entire aspects of the excitonic insulator states in the 
strong-coupling regime.  

\smallskip
\begin{acknowledgments}
We thank K. Okunishi for tutorial lectures, S. Yamamoto for enlightening discussion, 
C. E. Agrapidis for careful reading of our manuscript, and U. Nitzsche for technical 
assistance.  M.K. acknowledges the hospitality of IFW Dresden during his stay in 
Dresden and his use of computers.  
This work was supported in part by Grants-in-Aid for Scientific Research from 
JSPS (Projects No.~JP17K05530 and No.~JP19K14644), by the DFG through 
SFB 1143 (project-id 247310070), and by Keio University Academic Development 
Funds for Individual Research.  
\end{acknowledgments}


\begin{thebibliography}{99}
\bibitem{CMFT} There is a variety of the mean-field replacements:  
(i) several exchange bonds in the periodic finite-size cluster are replaced by the mean-field bonds \cite{albuquerque}, 
(ii) an open finite-size cluster is surrounded by the mean-field sites \cite{gotfryd}, and 
(iii) only one bond is replaced by the mean-field bond in the periodic finite-size chain \cite{shibata,singhania}, etc. 
Thus, there is some arbitrariness in the ways of introducing the mean-fields in CMFT. In a broad sense, 
the CMFT can be defined as a self-consistent determination of the mean-fields to minimize the 
total energy of a cluster where a part of sites and/or bonds are replaced by the mean-field sites 
and/or bonds.
\end{thebibliography}
\end{document}